\documentclass[twocolumn,showpacs,preprintnumbers,amsmath,amssymb,dvipdfmx]{revtex4}

\usepackage{graphicx}
\usepackage{dcolumn}
\usepackage{bm}
\usepackage{colortbl}
\usepackage{ulem}

\begin{document}

\title{
Field-Selective Anomaly and Chiral Mode Reversal in Type-II Weyl Materials
}

\author{M. Udagawa$^1$ and E.J. Bergholtz$^{2,3}$}%
\affiliation{%
$^1$Department of Physics, Gakushuin University, Mejiro, Toshima-ku, Tokyo 171-8588, Japan\\
$^2$Dahlem Center for Complex Quantum Systems and Institut f\"{u}r Theoretische Physik, Freie Universit\"{a}t Berlin, Arnimallee 14, 14195 Berlin, Germany\\
$^3$Department of Physics, Stockholm University, AlbaNova University Center, 106 91 Stockholm, Sweden }%

\date{\today}

\begin{abstract}
Three-dimensional condensed matter incarnations of Weyl fermions generically have a tilted dispersion---in sharp contrast with their elusive high-energy relatives where a tilt is forbidden by Lorentz invariance, and with the low-energy excitations of two-dimensional graphene sheets where a tilt is forbidden by either crystalline or particle-hole symmetry. Very recently, a number of materials (MoTe$_2$, LaAlGe and WTe$_2$) have been identified as hosts of so-called type-II Weyl fermions whose dispersion is so strongly tilted that a Fermi surface is formed, whereby the Weyl node becomes a singular point connecting electron and hole pockets. We here predict that these systems have remarkable properties in presence of magnetic fields. Most saliently, we show that the nature of the chiral anomaly depends crucially on the relative angle between the applied field and the tilt, and that an inversion-asymmetric overtilting creates an imbalance in the number of chiral modes with positive and negative slopes.
The field-selective anomaly gives a novel magneto-optical resonance, providing an experimental way to detect concealed Weyl nodes.
\end{abstract}

\maketitle

{\it Introduction.---} After eluding discovery for 85 years since their theoretical prediction \cite{weyl}, Weyl fermions---a fundamentally new type of massless particles \cite{volovikbook,murakami,wan,hosur}---were finally observed in 2015 \cite{weylexp,weylexp2}, and have by now been confirmed to exist as quasi-particles in a rapidly growing list of materials displaying remarkable properties including topological Fermi arc surface states \cite{wan,ojanen,weylexp,weylexp2} and a condensed matter incarnation of the chiral anomaly when exposed to external electromagnetic fields \cite{chiraltheo,chiralexp,chiralexp2,zhang}.

The Weyl Hamiltonian pertinent to condensed matter,  
\begin{eqnarray}
H_{{\rm Weyl}} = -{\mathbf v}^{0}\cdot{\mathbf k}\sigma^0 + \sum_{ij}v_{ij}k_i\sigma^j , 
\label{Effective_general_Hamiltonian}
\end{eqnarray}
where $\sigma^j$ are the Pauli matrices, encodes a generic linear crossing of two non-degenerate bands and notably includes a tilting term proportional to the unit matrix, $\sigma^0$ \cite{weylFCI,Max,soluyanov,comment}. In two dimensions, tilts may occur e.g. in organic conductors under pressure \cite{tilt2d} and in strained graphene \cite{tilt2dgraphene}. These effects are however quite small owing to the fragile nature of a band crossing point in two dimensions and the absence of tilts due to crystalline and/or particle-hole symmetries in these materials under ambient conditions. In three dimensions, in contrast, a non-degenerate band crossing is generic \cite{herring} and the presence of Weyl nodes are topological objects, monopole sources and drains of Berry flux, which can only be annihilated by merging Weyl nodes with opposite topological charge \cite{hosur}. This robustness make strongly tilted Weyl cones very natural, which give rises to unusual transport phenomena that are radically different compared to the effects of anisotropic velocities $v_{ij}$ \cite{Max}. 

As noted in Ref. \onlinecite{weylFCI}, three-dimensional Weyl cones may even be tilted over, implying the appearance of a finite Fermi surface yet keeping the topology entirely unchanged as a tilt does not influence the eigenstates of $H_{{\rm Weyl}}$. Ignited by the explicit materials prediction of overtilted (a.k.a. type-II) Weyl cones in the bandstucture of WTe$_2$ \cite{soluyanov}---a material that has also before this prediction received ample attention due to its remarkable transport properties, including a record breaking magnetoresistence \cite{titanic}---there is a present surge of experimental \cite{weyl2exp,weyl2exp2,weyl2exp3,weyl2exp4,weyl2exp5,weyl2expWTe2_1,weyl2expWTe2_2,weyl2expMoxW1?xTe2} and theoretical 
\cite{type2superfluid,type2prediction,aqhtype2,Koepernik,obrien,voloviktype2,tlitedDirac,autes,yu2016,minimal} studies of type-II Weyl systems. 

The first experimental identification type-II Weyl fermions were very recently reported in LaAlGe \cite{weyl2exp} and MoTe$_2$ \cite{weyl2exp2,weyl2exp3,weyl2exp4,weyl2exp5}, quickly followed by evidence for their presence also in WTe$_2$ \cite{weyl2expWTe2_1,weyl2expWTe2_2} and Mo$_x$W$_{1-x}$Te$_2$ \cite{weyl2expMoxW1?xTe2}. However, even a basic understanding of the possible relevance of the type-II Weyl fermions for the remarkable properties of these materials, especially in magnetotransport, is glaringly absent. Intuitively, it is not even clear whether the Weyl node should play any significant role in type-II materials as they are overwhelmed by the finite density of states of the hole and electron pockets attached to it. This motivates the present work in which we investigate the interplay of tilting the Weyl dispersion with a magnetic field, and thereby uncover several intriguing phenomena.

{\it Lattice model.---}
To examine how the overtilting affects the topological nature of a Weyl phase, we first consider a simple lattice model featuring Weyl nodes whose tilt is easily tuned:
\begin{align}
&H({\mathbf k}) = 2t[\sin k_x\sigma^x + \sin k_y\sigma^y + \cos k_z\sigma^z]\nonumber\\
 &+ 2m(2 - \cos k_x - \cos k_y)\sigma^z - 2t_{0x}\sin k_x\sigma^0.
\label{Lattice_Hamiltonian}
\end{align}
This Hamiltonian has a pair of Weyl nodes at ${\mathbf k} = {\mathbf k}_{\pm} = (0, 0, \pm\frac{\pi}{2})$. 
In fact, by linearizing this Hamiltonian around ${\mathbf k} = {\mathbf k}_{\pm}$, one obtains an effective Hamiltonian of the form (\ref{Effective_general_Hamiltonian}):
\begin{eqnarray}
H_{\rm eff}({\mathbf k}) = v(-\eta\delta k_x\sigma^0 + \sum_iw^{\pm}_i\delta k_i\sigma^i),
\label{effHamiltonian}
\end{eqnarray}
where $v=2t$, $\eta=t_{0x}/t$, and $\delta{\mathbf k}=(\delta k_x, \delta k_y, \delta k_z)={\mathbf k}-{\mathbf k}_{\pm}$.
$w^{\pm}_i = (1, 1, \mp1)$ for $i=x, y$ and $z$ defines the charge of Weyl node: $w_xw_yw_z=\mp1$.
Throughout this work we set $m/t=3$, $c=e=\hbar=1$, and fix the Fermi energy $\varepsilon_{\rm F}$ at the Weyl points: $\varepsilon_{\rm F}/t=0$, corresponding to the 
electron density at half filling. To make direct contact with experiments, we use $t=0.1$eV, and the lattice space, $a=5.0$\AA, which leads to the velocity, $v=1.0$eV\AA, comparable to the reported values for the archetypical Weyl material TaAs \cite{chiralexp}.
\begin{figure}[t]
\begin{center}
\includegraphics[width=0.45\textwidth]{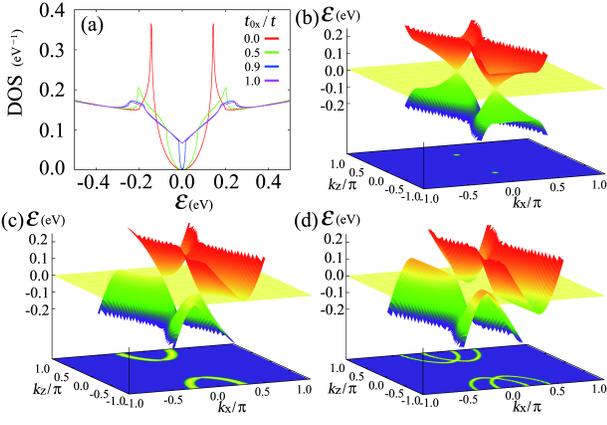}
\end{center}
\caption{\label{Fig1} 
(color online). (a) Density of states (DOS) of Hamiltonian (\ref{Lattice_Hamiltonian}) for $t_{0x}/t=0.0, 0.5, 0.9$ and $1.0$. (b), (c), (d) shows the $k_y=0$ projection of the band structure for $t_{0x}/t=0.0, 1.0, 1.5$. The Fermi levels are shown with yellow shaded planes. The projections of Fermi surface are also drawn at the bottom.
}
\end{figure}

At $t_{0x}/t=0.0$, low-energy excitations exist only around a pair of Weyl nodes resulting in semi-metallic density of states (DOS)  [Fig.~\ref{Fig1}(a) (b)]. In fact, the low-energy DOS is obtained as $D(\varepsilon)=\frac{\varepsilon^2}{16\pi^2t^3}$, giving the characteristic $\propto\varepsilon^2$ scaling law. 
As is clear from the linearlized Hamiltonian, eq.~(\ref{effHamiltonian}), the effect of tilting is introduced through $t_{0x}$, which breaks the inversion symmetry of the system, at the same time. In our model, this term originates from imaginary hopping, which occurs naturally e.g. in the context of double-exchange model with canted moments\cite{Ohgushi}, as can be seen in eq. (\ref{Lattice_Hamiltonian_Realspace}) at $B_z=B_x=0$.
As increasing $t_{0x}$, larger weight moves to low energy [Fig.~\ref{Fig1}(b) (c) (d)]. In particular, at $t_{0x}/t=1.0$ the generatrices of Weyl cones touch the Fermi level which causes a first-order Lifshitz transition and the appearance of finite DOS at $\varepsilon=\varepsilon_{\rm F}$.

\begin{figure*}[t]
\begin{center}
\includegraphics[width=0.8\textwidth]{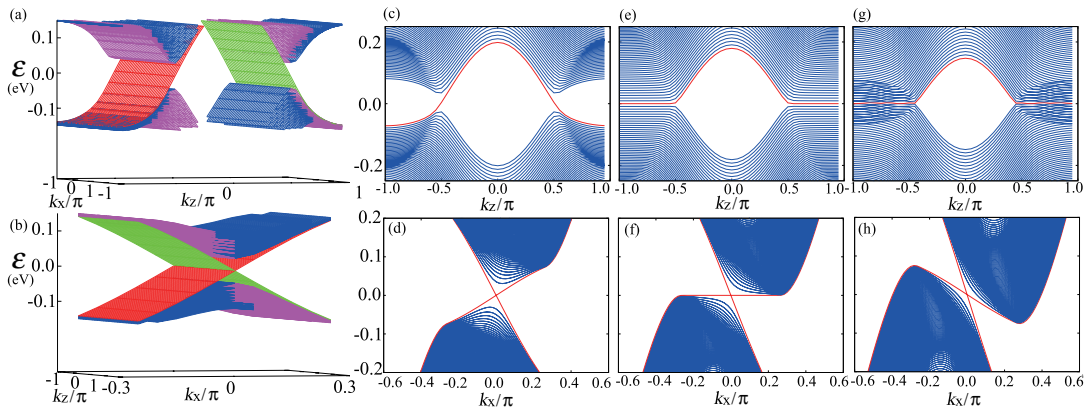}
\end{center}
\caption{\label{Fig2} 
(color online) Energy spectra are obtained for $L_y=400$, for (a), (c), (e), (g): ${\mathbf B}\parallel{\mathbf z}$ and (b), (d), (f), (h): ${\mathbf B}\parallel{\mathbf x}$. The values of $t_{0x}$ are chosen as (a), (b): $t_{0x}=0.0$, (c), (d): $0.5$, (e), (f): $1.0$, and (g), (h): $1.5$. The $L_y/2 + 1$-th modes, which may give chiral modes, are highlighted in red.
}
\end{figure*}

{\it Modified chiral anomaly.---} The chiral anomaly, associated with the appearance of chiral modes in presence of a magnetic field, is one of the most outstanding 
phenomena associated with Weyl semimetals. To examine how the tilting of Weyl nodes affects this property, we write the Hamiltonian (\ref{Lattice_Hamiltonian})
 in real space, and incorporate a magnetic field by Peierls substitution:
\begin{align}
H &= \frac{1}{i}\sum_{{\mathbf j},s,s'}\Bigl[(e^{iB_zj_y}c^{\dag}_{{\mathbf j}-{\mathbf x}s} - e^{-iB_zj_y}c^{\dag}_{{\mathbf j}+{\mathbf x}s})(\sigma^x_{ss'}-t_{0x}\sigma^0_{ss'})\nonumber\\ 
&+ 2im\Bigl\{2c^{\dag}_{{\mathbf j}s} - \frac{1}{2}(e^{iB_zj_y}c^{\dag}_{{\mathbf j}-{\mathbf x}s} + e^{-iB_zj_y}c^{\dag}_{{\mathbf j}+{\mathbf x}s})\nonumber\\
& - \frac{1}{2}(c^{\dag}_{{\mathbf j}-{\mathbf y}s} + c^{\dag}_{{\mathbf j}+{\mathbf y}s})\Bigr\}\sigma^z_{ss'} + (c^{\dag}_{{\mathbf j}-{\mathbf y}s} - c^{\dag}_{{\mathbf j}+{\mathbf y}s})\sigma^y_{ss'}\nonumber\\
&+ i(e^{-iB_xj_y}c^{\dag}_{{\mathbf j}-{\mathbf z}s} + e^{iB_xj_y}c^{\dag}_{{\mathbf j}+{\mathbf z}s})\sigma^z_{ss'}
\Bigr]c_{{\mathbf j}s'}
\label{Lattice_Hamiltonian_Realspace}
\end{align}
Here, ${\mathbf j}=(j_x, j_y, j_z)a$ is a coordinate on a cubic lattice.
We consider the magnetic field ${\mathbf B}=(B_x, 0, B_z)$ to be in the $x$-$z$ and choose vector potentials ${\mathbf A}({\mathbf r})=(-yB_z, 0, yB_x)$, which leads to the phase factors in eq.~(\ref{Lattice_Hamiltonian_Realspace}). 
We consider a system of dimension: $L_x\times L_y\times L_z$, and impose periodic boundary conditions for all the three directions.
To bring out the essential physics we will mainly focus on the two special cases of ${\mathbf B}\parallel{\mathbf x}$ or ${\mathbf B}\parallel{\mathbf z}$, and set the magnitude of magnetic field to $B=2\pi/L_y$.

We plot the energy spectra in Fig.~\ref{Fig2} for ${\mathbf B}\parallel{\mathbf z}$ and $\parallel{\mathbf x}$ for several values of $t_{0x}$. 
While the energies depend on two momenta, $k_x$ and $k_z$, the spectra are almost independent of the momentum perpendicular to magnetic field,
as is the case with a one-particle problem in a uniform magnetic field [Fig.~\ref{Fig2}(a), (b)].
At $t_{0x}/t=0.0$, one can clearly see the existence of a pair of chiral modes, one with positive, and the other with negative slope in terms of the
applied field direction. The sign of the slope is tightly connected with the Weyl charge of the original Weyl node.
As is clear from Fig.~\ref{Fig2}(a), the Weyl node with positive (negative) Weyl charge gives rise to a chiral mode with positive (negative) slope.
The conservation of Weyl charge inevitably leads to the appearance of both positive and negative chiral modes.

In Fig.~\ref{Fig2}(c)-(h), we plot the energy spectra in terms of the momentum parallel to magnetic field for increasing $t_{0x}$.
As we discuss below, the tilt of Weyl nodes due to finite $t_{0x}$ drastically changes the nature of chiral modes, depending on the applied magnetic field direction.
At $t_{0x}/t=0.5$, the chiral modes are clearly visible for both field directions [Fig.~\ref{Fig2}(c), (d)].
In the case of ${\mathbf B}\parallel{\mathbf x}$, one can see the velocities of chiral modes become inequivalent,
due to the inversion symmetry breaking by $t_{0x}/t$.

As increasing $t_{0x}$, a clear difference starts to show up in the evolution of chiral modes.
At $t_{0x}/t=1.0$, the Weyl nodes touch the Fermi level at $B=0$.
At this point, the chiral modes become indistinguishable from the bulk Landau levels, as shown in Fig.~\ref{Fig2}(e) for ${\mathbf B}\parallel{\mathbf z}$. 
Meanwhile, for ${\mathbf B}\parallel{\mathbf x}$, the chiral modes still remain visible, clearly separated from the bulk states. We note in passing that the the boundary between type-I and type-II Weyl behaviour, realised at $t_{0x}/t=1.0$ here, has been suggested as a stable phase in magnetic plasmas \cite{plasma}.

For $t_{0x}/t>1.0$, the Weyl cones are overtilted, and as a result the system cannot be regarded as a semimetal in the thermodynamic sense.
Nevertheless, depending on the magnetic field direction, the Weyl nodes can still play a crucial role. While for ${\mathbf B}\parallel{\mathbf z}$, chiral modes are missing, as shown in Fig.~\ref{Fig2}(g), for ${\mathbf B}\parallel{\mathbf x}$,
the two chiral modes remain clearly visible [Fig.~\ref{Fig2}(h)]. In the latter case, the persisting chiral modes show a peculiar feature: once the Weyl cones tilt over, the two chiral modes acquire the same sign for the slopes, as shown in Fig.~\ref{Fig2}(h). This is in sharp contrast to the semimetallic (type-I) regime where the conservation of Weyl charge implies that there is an equal number of chiral modes with positive and negative slopes. 

In order to understand this unusual behaviour, we focus on the linearized model (\ref{effHamiltonian}), and 
 investigate its magnetic response. We consider a single Weyl node, and incorporate the effect of magnetic field, by simply replacing 
 ${\mathbf k}$ with ${\mathbf k}-{\mathbf A}$.
Firstly, for ${\mathbf B} = (B, 0, 0)$, the spectrum for the positive (negative) Weyl node $E^{+(-)}_{n}$, is classified with Landau level index $n$, as
\begin{eqnarray}
E^{\pm}_{n} = -\eta vk_x+{\rm sgn}(n)\sqrt{2eB|n|v^2 + (vk_x)^2},
\end{eqnarray}
for $|n|\geq1$, and
\begin{eqnarray}
E^{\pm}_{0} = (\pm1-\eta)vk_x.
\label{zeromode_Bx}
\end{eqnarray}
for $n=0$. This $n=0$ mode corresponds to the chiral mode on which the particle propagates in only one direction, responsible for unusual magnetic response associated with chiral anomaly. As is clear from eq.~(\ref{zeromode_Bx}), the sign of velocity is firmly tied to the sign of Weyl charge
for untilted Weyl nodes ($\eta=0$). The tilting parameter $\eta$ reduces the velocity of one of the chiral modes, and eventually changes its sign at the critical value: $|\eta|=1$,
where the Weyl cone has tipped over.
The anomalous response expected from this chiral mode, e.g. charge pumping, is expected to change its sign, accordingly.
 
The analysis can be extended to a general field direction.
With ${\mathbf B} = (B\cos\theta, 0, B\sin\theta)$ the energy spectrum is
\begin{align}
E_n^{\pm} = -vk_{\parallel}&\eta\cos\theta+{\rm sgn}(n)\sqrt{1-\eta^2\sin^2\theta}\nonumber\\
&\times\Bigl[(vk_{\parallel})^2 + 2eB|n|v^2\sqrt{1-\eta^2\sin^2\theta}\Bigr]^{\frac{1}{2}}
\label{Tiltcone_Landau_generaltheta}
\end{align}
for $|n|\geq1$, and 
\begin{align}
E_0^{\pm} = vk_{\parallel}(-\eta\cos\theta\pm\sqrt{1-\eta^2\sin^2\theta})
\label{zeromode_generaltheta}
\end{align}
for $n=0$, with $k_{\parallel}$, the momentum parallel to magnetic field.
These solutions are, however, valid only for the field nearly parallel to ${\mathbf x}$: $1-\eta^2\sin^2\theta>0$. Outside this region, the effective linearized model breaks down and the chiral modes disappear, consistent with the lattice model [Fig.~\ref{Fig2}(g), (h)], i.e. the chiral anomaly appears only for a limited field direction. 
The validity of the solution depends on the topology of semiclassical cyclotron orbit. Only if the above condition: $1-\eta^2\sin^2\theta>0$ is satisfied, the orbit closes, encircling the Weyl nodes, enabling the description of low-energy states using the linearized model.

\begin{figure}[t]
\begin{center}
\includegraphics[width=0.49\textwidth]{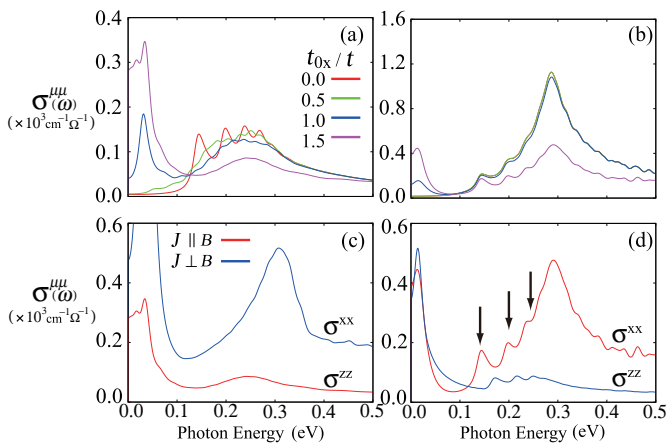}
\end{center}
\caption{\label{Fig3} 
(color online) (a) (b): The optical conductivities parallel to ${\mathbf B}$ are plotted for several $t_{0x}$'s. (a) $\sigma^{zz}(\omega)$ for ${\mathbf B}\parallel{\mathbf z}$ and (b) $\sigma^{xx}(\omega)$ for ${\mathbf B}\parallel{\mathbf x}$ are plotted. (c) (d): The optical conductivities in the overtilted regime: $t_{0x}/t=1.5$. $\sigma^{zz}(\omega)$ and $\sigma^{xx}(\omega)$ are plotted for (c) ${\mathbf B}\parallel{\mathbf z}$ and (d) ${\mathbf B}\parallel{\mathbf x}$. The arrows indicate the transition energies between $n$-th and $-n$-th Landau levels for $n=1, 2$ and $3$ from the lower side.
}
\end{figure}

{\it Optical conductivity.---} The field-angle dependence of Landau levels (\ref{Tiltcone_Landau_generaltheta}) and associated chiral modes (\ref{zeromode_generaltheta}) implies that a sensitive experimental probe is available for the detection of overtilted Weyl nodes in a magnetic field. 
For untilted Weyl nodes, the characteristic power-law behavior of thermodynamic quantities provides straightforwardly measurable experimental signatures.
In contrast, for the overtilted Weyl nodes, the existence of a Fermi surface masks the contribution from the Weyl nodes.
A possible avenue to detect the concealed Weyl nodes may be to look at a transition spectrum from the filled Fermi sea.
We suggest that the magneto-optical conductivity can serve as a convenient probe in this context.

The optical conductivity $\sigma^{\mu\mu}$ is obtained by the Kubo formula:
\begin{eqnarray}
\sigma^{\mu\mu}(\omega) = \frac{1}{iV}\sum_{\alpha,\alpha'}\frac{f_{\alpha} - f_{\alpha'}}{\varepsilon_{\alpha} - \varepsilon_{\alpha'}}\frac{\langle\alpha|J_{\mu}|\alpha'\rangle\langle\alpha'|J_{\mu}|\alpha\rangle}{\omega + \frac{i}{\tau} - (\varepsilon_{\alpha} - \varepsilon_{\alpha'})},
\label{Optical}
\end{eqnarray}
where $\varepsilon_{\alpha}$ and $|\alpha\rangle$ are the eigenenergy and eigenstate of the Hamiltonian (\ref{Lattice_Hamiltonian_Realspace}), and
$J_{\mu}\equiv\frac{1}{i}[\sum_{js}j_{\mu}c^{\dag}_{js}c_{js}, H]$ gives the charge current operator.
As a small damping factor, we have introduced $\frac{1}{\tau}=0.1t$ to mimic impurity scattering, and the temperature is set to be $T=0.01t$.
This choice of parameters corresponds to $T\sim$10K and the mean free path, $l\sim100$\AA, well within the reach of experiments.
The $\sigma^{\mu\mu}(\omega)$ contains both contributions from the Weyl nodes and the Fermi surface. We scrutinize the spectrum, and find
the robust feature attributed to Weyl nodes, which survive the dominant contribution from the Fermi surface.

In Fig.~\ref{Fig3} (a) and (b), we plot the $t_{0x}$ dependence of $\sigma^{\mu\mu}(\omega)$ for both ${\mathbf B}\parallel{\mathbf z}$ and ${\mathbf B}\parallel{\mathbf x}$. For both field directions, $\sigma^{\mu\mu}(\omega)$ commonly develops a peak in the low-energy regions around $t_{0x}/t=1.0$, where the Weyl node is completely tilted. These peaks are attributed to the low-energy particle-hole excitations near the Fermi surface. 

Meanwhile, the spectra in higher energy part shows considerable difference. For ${\mathbf B}\parallel{\mathbf z}$, a clear magneto-oscillation can be found for $t_{0x}/t=0.0$. Whereas, the oscillation becomes quickly damped as $t_{0x}$ increases.  
In contrast, for ${\mathbf B}\parallel{\mathbf x}$, a clear quantum oscillation persists as $t_{0x}$ increases, even after the Weyl nodes are completely tilted [Fig.~\ref{Fig3} (a) and (b)]. 

We plot optical conductivities both parallel and perpendicular to magnetic field in Fig.~\ref{Fig3} (c) and (d).
Clear quantum oscillation can be seen for both components for ${\mathbf B}\parallel{\mathbf x}$, while the oscillations are completely smoothed out for ${\mathbf B}\parallel{\mathbf z}$.
The difference in optical response can be attributed to the field-selective quantum anomaly, discussed above.
For ${\mathbf B}\parallel{\mathbf x}$, the Landau levels originate from the cyclotron orbit around the Weyl points, resulting in a clear indirect gap ($\propto \sqrt{eB}v$), which separates the $n>0$ and $n<0$ parts of the spectrum, traversed by the $n=0$ chiral modes [Fig.~\ref{Fig2} (h)].
This level structure results in robust peaks in the optical spectra due to the transitions between $n$-th and $-n$-th Landau levels, as 
indicated with black arrows in Fig.~\ref{Fig3} (d). 
In other words, this resonance structure is robust against carrier doping as long as the occupancy on the low-index Landau levels is moderate, which also explains why the optical resonance is clearly visible in spite of the masking by low-energy continuum due to the 
particle-hole excitations around the Fermi surface.
In contrast, for ${\mathbf B}\parallel{\mathbf z}$, the Landau levels form almost continuous structure [Fig.~\ref{Fig2} (g)], which results in the absence of clear resonance peaks [Fig.~\ref{Fig3} (c)].
This field-orientational optical resonance serves as a useful diagnose to experimentally detect the elusive overtilted Weyl nodes.

{\it Summary.---}  We have studied the effect of tilting Weyl nodes on the chiral anomaly with a simple lattice model with possible relevance to a number of materials that are currently intensively studied theoretically as well as experimentally. In the type-II regime of overtilted Weyl cones, realised in e.g. MoTe$_2$, LaAlGe and WTe$_2$, we found that a magnetic field can reinforce the importance of the Weyl nodes depending on the angle between the tilt and the applied field despite the fact that the Weyl node is shadowed by a finite Fermi surface. We furthermore found that the number of chiral modes with positive and negative slopes can be different in systems with overtilted Weyl nodes, in striking contrast to
the conservation of Weyl charges for untilted Weyl nodes.
We clarified its origin 
through the effective linearized Hamiltonian. We further proposed the optical conductivity, and signatures of quantum oscillations therein, as a useful diagnostic to detect the overtilted Weyl nodes in the presence of formidable screening from Fermi surface. 

{\it Acknowledgements.---} We thank M. Trescher for discussions. This work was supported by JSPS KAKENHI (Nos. 26400339, 15H05852, 15K13533 and 16H04026), 
DFG's Emmy Noether program (BE 5233/1-1), and the Wallenberg Academy Fellows program. 
{\it Note added:} While finalising this manuscript, Ref. \onlinecite{yu2016}, which has some overlap with our work, occurred on the arXiv, although it entirely focuses on the effects of a magnetic field on a single Weyl cone in contrast to our work.


\begin{thebibliography}{9}
%
\bibitem{weyl} H. Weyl, Z. Physik {\bf 56}, 330 (1929).
\bibitem{volovikbook} G. E. Volovik, The Universe in a Helium Droplet (Oxford University Press (OUP), 2009).
\bibitem{murakami} S. Murakami, New Journal of Physics {\bf 9}, 356 (2007).
\bibitem{wan} X. Wan, A. M. Turner, A. Vishwanath, and S. Y. Savrasov, Phys. Rev. B {\bf 83}, 205101 (2011).

\bibitem{hosur}P. Hosur and X. Qi, C. R. Phys. {\bf 14}, 857 (2013).

\bibitem{weylexp} S.-Y. Xu, I. Belopolski, N. Alidoust, M. Neupane, G. Bian, C. Zhang, R. Sankar, G. Chang, Z. Yuan, C.-C. Lee, S.-M. Huang, H. Zheng,
J. Ma, D. S. Sanchez, B. Wang, A. Bansil, F. Chou, P. P. Shibayev, H. Lin, S. Jia, and M. Z. Hasan, Science {\bf 349}, 613 (2015).
\bibitem{weylexp2} B. Q. Lv, H. M. Weng, B. B. Fu, X. P. Wang, H. Miao, J. Ma, P. Richard, X. C. Huang, L. X. Zhao, G. F. Chen, Z. Fang, X. Dai, T. Qian,
and H. Ding, Phys. Rev. X {\bf 5}, 031013 (2015).

\bibitem{ojanen}T. Ojanen, Phys. Rev. B {\bf 87}, 245112 (2013).

\bibitem{chiraltheo}H. B. Nielsen and M. Ninomiya, Physics Letters B, {\bf 130}, 389 (1983); V. Aji, Phys. Rev. B {\bf 85}, 241101 (2012); C.-X. Liu, P. Ye, and X.-L. Qi,
Phys. Rev. B {\bf 87}, 235306 (2013); D.T. Son and N. Yamamoto,
Rev. Lett. {\bf 109}, 181602 (2012);  A.G. Grushin, Phys. Rev. D {\bf 86}, 045001 (2012);
A. A. Zyuzin and A. A. Burkov, Phys. Rev. B {\bf 86}, 115133 (2012); P. Goswami and S. Tewari, Phys. Rev. B {\bf 88}, 245107 (2013); S.A. Parameswaran, T. Grover, D.A. Abanin, D.A. Pesin, and A. Vishwanath, Phys. Rev. X {\bf 4}, 031035 (2014); J. Behrends, A. G. Grushin, T. Ojanen, and J. H. Bardarson,
Phys. Rev. B {\bf 93}, 075114 (2016).

\bibitem{chiralexp} X. Huang, L. Zhao, Y. Long, P. Wang, D. Chen, Z. Yang, H. Liang, M. Xue, H. Weng, Z. Fang, X. Dai, and G. Chen, Phys. Rev. X {\bf 5}, 031023 (2015).

\bibitem{chiralexp2} C. Zhang, S.-Y. Xu, I. Belopolski, Z. Yuan, Z. Lin, B. Tong, N. Alidoust, C.-C. Lee, S.-M. Huang, H. Lin, M. Neupane, D. S. Sanchez, H. Zheng, G. Bian, J. Wang, C. Zhang, T. Neupert, M. Z. Hasan, and S. Jia, arXiv:1503.02630.
%

\bibitem{zhang} C.-L. Zhang, S.-Y. Xu, I. Belopolski, Z. Yuan, Z. Lin, B. Tong, G. Bian, N. Alidoust, C.-C. Lee, S.-M. Huang, T.-R. Chang,	G. Chang,	 C.-H. Hsu, H.-T. Jeng, M. Neupane,	D.S. Sanchez,	H. Zheng, J. Wang, H. Lin, C. Zhang, H.-Z. Lu, S.-Q Shen, T. Neupert, M. Z. Hasan, and S. Jia, Nat. Comms. {\bf 7},
10735 (2016).

\bibitem{weylFCI} E.J. Bergholtz, Z. Liu, M. Trescher, R. Moessner, and M. Udagawa, Phys. Rev. Lett. {\bf 114}, 016806 (2015).
\bibitem{Max} M. Trescher, B. Sbierski, P. W. Brouwer, and E. J. Bergholtz, Phys. Rev. B {\bf 91}, 115135 (2015).
\bibitem{soluyanov} A. A. Soluyanov, D. Gresch, Z. Wang, Q. Wu, M. Troyer, X. Dai, and B. A. Bernevig, Nature {\bf 527}, 495 (2015).
\bibitem{comment} C. Beenakker, Journal Club for Condensed Matter Physics, August 2015 (2015).

\bibitem{tilt2d} A. Kobayashi, S. Katayama, Y. Suzumura, and H. Fukuyama, J. Phys. Soc. Jpn. {\bf 76}, 034711 (2007).

\bibitem{tilt2dgraphene} M.O. Goerbig, J.-N. Fuchs, G. Montambaux, F. Piechon, Phys. Rev. B {\bf 78}, 045415 (2008).

\bibitem{herring} C. Herring, Phys. Rev. {\bf 52} 365 (1937).

\bibitem{titanic}  M. N. Ali,	J. Xiong, S. Flynn, J. Tao,	Q. D. Gibson, L. M. Schoop, T. Liang, N. Haldolaarachchige, M. Hirschberger, N. P. Ong, and R. J. Cava, Nature {\bf 514}, 205 (2014).

\bibitem{weyl2exp} S.-Y. Xu, N. Alidoust, G. Chang, H. Lu, B. Singh, I. Belopolski, D. Sanchez, X. Zhang, G. Bian, H. Zheng, M.-A. Husanu, Y. Bian, S.-M.
Huang, C.-H. Hsu, T.-R. Chang, H.-T. Jeng, A. Bansil, V. N. Strocov, H. Lin, S. Jia, and M. Z. Hasan, arXiv:1603.07318v3.

\bibitem{weyl2exp2} K. Deng, G. Wan, P. Deng, K. Zhang, S. Ding, E. Wang, M. Yan, H. Huang, H. Zhang, Z. Xu, J. Denlinger, A. Fedorov, H. Yang, W. Duan,H. Yao, Y. Wu, y. S. Fan, H. Zhang, X. Chen, and S. Zhou, arXiv:1603.08508v1.

\bibitem{weyl2exp3} L. Huang, T. M. McCormick, M. Ochi, Z. Zhao, M.-t. Suzuki, R. Arita, Y. Wu, D. Mou, H. Cao, J. Yan, N. Trivedi, and A. Kaminski,
arXiv:1603.06482v1.

\bibitem{weyl2exp4} N. Xu, Z. J. Wang, A. P. Weber, A. Magrez, P. Bugnon, H. Berger, C. E. Matt, J. Z. Ma, B. B. Fu, B. Q. Lv, N. C. Plumb, M. Radovic, E. Pomjakushina, K. Conder, T. Qian, J. H. Dil, J. Mesot, H. Ding, and M. Shi, arXiv:1604.02116.

\bibitem{weyl2exp5}A. Liang, J. Huang, S. Nie, Y. Ding, Q. Gao, C. Hu, S. He, Y. Zhang, C. Wang, B. Shen, J. Liu, P. Ai, Li Yu, X. Sun, W. Zhao, S. Lv, D. Liu, C. Li, Y. Zhang, Y. Hu, Y. Xu, L. Zhao, G. Liu, Z. Mao, X. Jia, F. Zhang, S. Zhang, F. Yang, Z. Wang, Q. Peng, H. Weng, X. Dai, Z. Fang, Z. Xu, C. Chen, and X. J. Zhou, arXiv:1604.01706.

\bibitem{weyl2expWTe2_1}C. Wang, Y. Zhang, J. Huang, S. Nie, G. Liu, A. Liang, Y. Zhang, B. Shen, J. Liu, C. Hu, Y. Ding, D. Liu, Y. Hu, S. He, L. Zhao, L. Yu, J. Hu, J. Wei, Z. Mao, Y. Shi, X. Jia, F. Zhang, S. Zhang, F. Yang, Z. Wang, Q. Peng, H. Weng, X. Dai, Z. Fang, Z. Xu, C. Chen, X. J. Zhou, arXiv:1604.04218.

\bibitem{weyl2expWTe2_2}Y. Wu, N.H. Jo, D. Mou, L. Huang, S. L. Bud'ko, P. C. Canfield, and A. Kaminski,  arXiv:1604.05176  

\bibitem{weyl2expMoxW1?xTe2}I. Belopolski, S.-Y. Xu, Y. Ishida, X. Pan, P. Yu, D. S. Sanchez, M. Neupane, N. Alidoust, G. Chang, T.-R. Chang, Y. Wu, G. Bian, H. Zheng, S.-M. Huang, C.-C. Lee, D. Mou, L. Huang, Y. Song, B. Wang, G. Wang, Y.-W. Yeh, N. Yao, J. E. Rault, P. Le Fevre, F. Bertran, H.-T. Jeng, T. Kondo, A. Kaminski, H. Lin, Z. Liu, F. Song, S. Shin, M. Z. Hasan, arXiv:1604.07079.

\bibitem{type2superfluid}Y. Xu, F. Zhang and C. Zhang, Phys. Rev. Lett. {\bf 115}, 265304 (2015).

\bibitem{type2prediction}T.-R. Chang, S.-Y. Xu, G. Chang, C.-C. Lee, S.-M. Huang, B.K. Wang, G. Bian, H. Zheng, D. S. Sanchez, I. Belopolski, N. Alidoust, M. Neupane, A. Bansil, H.-T. Jeng, H. Lin, and M. Z. Hasan, Nat. Commun. {\bf 7} 10639 (2016).

\bibitem{aqhtype2}A. A. Zyuzin and R. P. Tiwari, arXiv:1601.00890.

\bibitem{Koepernik} K. Koepernik, D. Kasinathan, D.V. Efremov, S. Khim, S. Borisenko, B. B\"uchner, and J. van den Brink, arXiv:1603.04323.

\bibitem{obrien} T. E. O'Brien, M. Diez, C. W. J. Beenakker, arXiv:1604.01028.

\bibitem{voloviktype2} G.E. Volovik, arXiv:1604.00849.

\bibitem{tlitedDirac}L. Muechler, A. Alexandradinata, T. Neupert, and R. Car, arXiv:1604.01398.

\bibitem{autes}G. Autes, D. Gresch, A. A. Soluyanov, M. Troyer, and O. V. Yazyev, arXiv:1603.04624.

\bibitem{yu2016}Z. Yu, Y. Yao, and S.A. Yang, arXiv:1604.04030. 

\bibitem{minimal}T. M. McCormick, I. Kimchi, and N. Trivedi, arXiv:1604.03096. 

\bibitem{Ohgushi} K. Ohgushi, S. Murakami and N. Nagaosa, Phys. Rev. B {\bf 62}, R6065 (2012).

\bibitem{plasma}W. Gao, B. Yang, M. Lawrence, F. Fang, B. Beri, and S. Zhang, arXiv:1511.04875.


\end{thebibliography}
\end{document}